\documentclass[11pt]{article}
\usepackage{latexsym,cite,amssymb}
\usepackage{times}
\makeatletter

\@addtoreset{equation}{section} \makeatother

\input amssym.def
\input amssym.tex

\newcommand{\barchi}{{\bar{\chi}}}
\newcommand{\barve}{{\bar{\varepsilon}}}
\newcommand{\bareta}{{\bar{\eta}}}

\newcommand{\dalpha}{{\dot{\alpha}}}
\newcommand{\dbeta}{{\dot{\beta}}}

\newcommand{\G}{{\Gamma}}
\newcommand{\g}{{\gamma}}

\newcommand{\half}{{{\textstyle\frac{1}{2}}}}

\newcommand{\be}{\begin{equation} }
\newcommand{\ee}{\end{equation} }
\newcommand{\ba}{\begin{array}}
\newcommand{\ea}{\end{array}}

\newcommand{\so}{\mbox{so}}
\newcommand{\SO}{\mbox{SO}}

\def\cC{{\cal C}}
\def\cD{{\cal D}}
\def\cE{{\cal E}}
\def\cL{{\cal L}}

\def\cD{{\cal D}}

\def\cS{{\cal S}}

\def\cZ{{\cal Z}}

\def\Tr{{\rm Tr}}
\def\I_N{{1_{\scriptscriptstyle N\times N}}}

\def\invsg{{\textstyle{\frac{1}{\sqrt{g}\,}}}}

\def\cN{{\cal  N}}

\newcommand{\p}{\partial}

\def\a{\alpha}
\def\b{\beta}
\def\g{\gamma}
\def\G{\Gamma}
\def\d{\delta}

\def\l{\lambda}
\def\L{\Lambda}
\def\m{\mu}
\def\n{\nu}

\def\s{\sigma}
\def\S{\Sigma}

\newcommand{\bea}{\begin{eqnarray}}
\newcommand{\eea}{\end{eqnarray}}
\newcommand{\bn}{\begin{enumerate}}
\newcommand{\en}{\end{enumerate}}

\def\CA{{\cal A}}

\def\CL{{\cal L}}

\def\CN{{\cal N}}
\def\CO{{\cal O}}

\def\CS{{\cal S}}

\def\a{\alpha}
\def\b{\beta}
\def\g{\gamma}
\def\d{\delta}
\def\e{\epsilon}
\def\ve{\varepsilon}

\def\l{\lambda}
\def\m{\mu}
\def\n{\nu}

\def\s{\sigma}

\def\G{\Gamma}

\def\L{\Lambda}
\def\S{\Sigma}

\def\hl{\hat{\lambda}}
\def\hm{\hat{\mu}}
\def\hn{\hat{\nu}}

\def\goto{\rightarrow}

\def\p{\partial}

\def\Tr{{\rm Tr}}

\def\det{{\rm det}}

\def\da{{\dot{\a}}}
\def\db{{\dot{\b}}}

\begin{document}
\begin{titlepage}
\title{
\vskip 2cm
Topological Twisting of Multiple M2-brane Theory\\}
\author{\sc
Kanghoon Lee,${}^{\sharp}~$ Sangmin Lee${}^{\ast}~$ ~and~ Jeong-Hyuck Park${}^{\dagger}$}
\date{}
\maketitle \vspace{-1.0cm}
\begin{center}
~~~\\
${}^{\sharp}$Department of Physics, College of Science, Yonsei University, Seoul 120-749, Korea
~{}\\
${}^{\ast}$Department of Physics and Astronomy, Seoul National University, Seoul 151-747, Korea
~{}\\
${}^{\dagger}$Department of Physics, Sogang University, Seoul 121-742, Korea\\
~{}\\
~~~\\
~~~\\
\end{center}
\begin{abstract}
\vskip0.5cm
\noindent
Bagger-Lambert-Gustavsson theory with infinite dimensional gauge group has been  suggested to describe M5-brane as a  condensation of multiple  M2-branes. Here we perform a topological twisting of  the Bagger-Lambert-Gustavsson theory.  The original    $\SO(8)$ $R$-symmetry is broken to   $\SO(3)\times \SO(5)$, where the former may be interpreted  as a diagonal subgroup of the Euclidean M5-brane world-volume symmetry $\SO(6)$, while the latter is the isometry of the transverse five directions. Accordingly
the resulting action contains  an one-form  and five scalars  as for the   bosonic dynamical  fields.  We further lift the  action to a generic curved three manifold.
In order to make sure the genuine topological invariance, we construct an off-shell supersymmetric formalism such that the scalar supersymmetry transformations  are nilpotent strictly off-shell and independent of the metric of the three manifold.
The one loop partition function around a trivial background
yields the Ray-Singer torsion. The BPS equation involves an  M2-brane charge density given by a Nambu-Goto action defined in an internal three-manifold.
\end{abstract}
{\small
\begin{flushleft}
\end{flushleft}}
\end{titlepage}
\newpage

\tableofcontents 
\section{Introduction}
D-branes have played a crucial role in understanding
non-perturbative dynamics of string theory.
The M2 and M5 branes
are expected to play a similar role in ${\cal{M}}$-theory, but due to their intrinsically non-perturbative nature, their world-volume theories remain much less understood than those of D-branes.
In particular, a Lorentz invariant
Lagrangian descriptions of the interacting conformal field theories living in the M2/M5 world-volume have been missing.

Being the only branes in ${\cal M}$-theory (in flat eleven-dimensions),
M2 and M5 branes are intricately related.
First, they are electromagnetic dual to each other
with respect to the four-form field strength
$G_{(4)}={\rm d}C_{(3)}$ in the eleven-dimensional supergravity.
Second, M2-branes can end on M5-branes just as
fundamental strings end on D-branes.
Roughly speaking, quantum excitations of open M2-branes should give rise
to a microscopic formulation of the M5-brane world-volume theory.
Third, the self-dual three-form flux $H_{(3)}$ on M5-branes carries
M2-brane charge. Finally, M2-branes in a background $G_{(7)}=*G_{(4)}$
can be blown up to M5-branes by an ${\cal M}$-theory version of the Myers' effect \cite{Myers:1999ps}.

Some time ago, Basu and Harvey \cite{Basu:2004ed} studied
the BPS configuration of M2-branes ending on M5-branes,
which exhibits many of these relations at once.\footnote{See Ref.\cite{Bonelli:2008kh} for further discussion.}
In analogy with the D1-D3 interpretation of Nahm's equations \cite{Diaconescu:1996rk, Tsimpis:1998zh},
they argued that the ``non-Abelian'' M2-brane world-volume theory
should admit a sort of fuzzy three-sphere solution \cite{Guralnik:2000pb}.
Around the same time, from a different perspective, Schwarz \cite{Schwarz:2004yj} raised the possibility of using superconformal Chern-Simons theories as for the description of
the  M2-brane dynamics.

Inspired by these pioneering works, Bagger-Lambert \cite{BL}
and Gustavsson \cite{Gus} (BLG) succeeded in writing down
an $\CN=8$ superconformal Chern-Simons-matter theory with $\SO(8)$ $R$-symmetry. 
The BLG Lagrangian was interpreted as the low energy of limit of the 
world-volume theory of two M2-branes in a certain M-theory background 
\cite{Lambert:2008et,Distler:2008mk}.

The action is based on a  gauge symmetry generated by the  so-called three-algebra. As for a conventional, ghost-free field theory with a \textit{finite} number of fields, the BLG theory admits only one gauge group
$\SO(4) \simeq \mbox{SU}(2)\times \mbox{SU}(2)$  with opposite levels for the two Chern-Simons terms, and matter
fields come in bi-fundamental representations. The uniqueness
is due to the surprisingly strong constraint imposed by the  three-algebra structure.
In order to free this severe restriction, one can consider either Lorentzian gauge groups~\cite{lorentzian} or infinite dimensional three-algebras. The latter can be realized as a volume-preserving diffeomorphism of an auxiliary
three-dimensional manifold. Combining both the original and the auxiliary  three manifolds leads to a six-dimensional manifold, and the BLG theory with  an infinite-dimensional gauge group may  have a natural origin as an
M5-brane action~\cite{Ho:2008nn,Park:2008qe,Bandos:2008fr}. In particular, in Ref.\cite{Park:2008qe} it has been shown that by generalizing the Brink-Di Vecchia-Howe-Polyakov method, Nambu-Goto action for a $p$-brane can be reformulated as a
$d$-dimensional gauged nonlinear sigma model  having a Nambu $(p+1-d)$-bracket squared potential. While the choice $d=p-1$ leads to the Yang-Mills potential, the choice $d=p-2$ leads to the Nambu three-bracket potential, and hence an infinite dimensional three-algebra. In particular, an M5-brane may be described by a condensation of M2-branes.

The connection between multiple M2's and an M5 motivates us to twist the (Euclidean) BLG theory by diagonalizing the $\SO(3)$ Lorentz symmetry and an $\SO(3)$ subgroup of the $\SO(8)$ $R$-symmetry.  The resulting action will contain  five scalars which can be viewed as the physical degrees of freedom along the five transverse directions of an M5-brane. While the twisting we perform works for any three-algebra, as an application, we will consider infinite dimensional gauge group or volume-preserving diffeomorphism in an internal three manifold at the end of the paper.

A quantum field theory is called topological if all vacuum
expectation values (vevs) of a certain set of operators (¡®observables¡¯) are metric-independent.
In particular, topological quantum field theories (TQFTs) of cohomological type are constructed as follows. Let us assume there is
a nilpotent symmetry of the action $Q$, such that $Q^2 = 0$. It follows that, at least formally, one
can deform the Lagrangian by adding an arbitrary $Q$-exact term without affecting the partition
function or the vevs of  observables (which are defined as elements in the cohomology of $Q$).
Since $Q$ is a symmetry of the action, the Lagrangian can be expressed as a sum of a $Q$-exact
and a $Q$-closed piece. The theory is therefore independent of any coupling constant in the $Q$-exact
piece. Moreover, if the energy momentum tensor is $Q$-exact all vevs of observables are
metric-independent and the theory is topological.

The organization of the present paper is as follows.

In section \ref{seconshell}, we construct the twisted BLG theory.
We begin with writing down the Euclidean version
of the BLG theory. Then, we perform a twist
which preserves an $\SO(3)\times \SO(5) \subset \SO(8)$ $R$-symmetry.
On-shell nilpotency  of the scalar  supersymmetries   and
the corresponding  BPS equations are also presented.

In section \ref{secoffshell}, to make sure the genuine topological invariance,
we introduce some auxiliary fields such that
the supersymmetry algebra closes strictly off-shell and the supersymmetry transformations are independent of the three-manifold metric.
Using the off-shell supersymmetric formulation,
we separate the twisted BLG action into  a $Q$-closed topological part and a $Q$-exact part,
thereby verifying  the topological invariance of the theory.

In section \ref{secobs}, we initiate
the study of observables of the theory.
We explicitly derive those observables which can be obtained
from the Lagrangian through a descent relation.
Then, we explore the possibility of a Wilson-loop operator,
but our analysis indicates that the twisted BLG theory does not admit a $Q$-closed Wilson loop operator.
Then we take a first step toward the perturbative
computation of the partition function.
The one loop determinants around a trivial background
turns out to be the Ray-Singer torsion.

In section \ref{secM5}, we interpret  our results from the   M5-brane point of view.  Realizing infinite dimensional gauge symmetry as volume preserving diffeomorphism in an internal three manifold, our twisted  theory can be viewed as partial topological twisting of a six-dimensional theory, where the  six-dimensional space has the fiber bundle structure: at each point in a three manifold (base), there exists a corresponding  internal three manifold  (fiber). The BPS equations then  involves an  M2-brane charge density given by a Nambu-Goto action defined in an internal three-manifold.

In section \ref{seccon}, we conclude  with some comments on future work.

Appendix carries some relevant useful identities.


\section{Twisted Bagger-Lambert-Gustavsson theory\label{seconshell}}

\subsection{Euclidean Bagger-Lambert-Gustavsson theory}
To start, we present the Euclidean version of the Bagger-Lambert-Gustavsson  Lagrangian:
\be
\ba{ll}
\CL_{\rm Euclidean} = &i \e^{\m\n\l} \left(\half f^{abcd} A_{\m ab} \p_\n A_{\l cd}  -\frac{1}{3} f^{cdag} f^{efb}{}_g A_{\m ab} A_{\n cd} A_{\l ef} \right)\\
&  +\Tr\Big[ \half (D_\m X^I)^2 - \frac{i}{2} \bar{\Psi} \G^\m D_\m \Psi
+\frac{i}{4}\bar{\Psi} \G_{IJ}[X^I,X^J,\Psi] + \frac{1}{12} [X^I, X^J, X^K]^2\Big]\,.
\ea
\label{EuclideanL}
\ee
There are some common as well as distinct  features compared to the original  Minkowskian case~\cite{BL}.
In terms of an explicit  basis of  the three-algebra,
\be
[T^{a},T^{b},T^{c}]=f^{abc}{}_{d}T^{d}\,,
\ee
the dynamical variables take values in the three-algebra, \textit{e.g.} $X^{I}=X^{I}_{a}T^{a}$. The trace is always taken over second-order in  three-algebra variables such that, in fact it involves a metric which can  raise or low the gauge index. The covariant derivatives are the same as in the Minkowskian case~\cite{BL}:
\be
\ba{ll}
D_\m X_a^I = \p_\m X_a^I +\tilde{A}_{\m a}{}^b X_b^I\,,~~~~&~~~~
D_\m X^{aI} = \p_\m X^{aI} - X^{bI}\tilde{A}_{\m b}{}^{a}= \p_\m X^{aI} +\tilde{A}_{\m}{}^{a}{}_{b} X^{bI}\,.
\ea
\ee
The tilde symbol  denotes the contraction with the  structure constant of the three-algebra,
\be
\tilde{A}_{\m a}{}^b:=A_{\m cd}f^{cd}{}_{a}{}^{b}\,.
\ee
The gauge symmetry is then realized by
\be
\ba{lll}
\d_\L X_a^I = -\tilde{\L}_a{}^b X^I_b\,,~~&~~
\d_\L \Psi_a = -\tilde{\L}_a{}^b  \Psi_b\,,~~&~~
\d_\L A_{\m ab} = D_\m \L_{ab}=\p_\m \L_{ab} + \tilde{A}_{\m a}{}^c \L_{cb} +\tilde{A}_{\m b}{}^c \L_{ac}\,.
\ea
\label{gaugetr}
\ee
The key difference, compared to the Minkowskian signature~\cite{BL},  is that  the Euclidean  action contains only the `holomorphic' part of the spinor such that $\bar{\Psi}$ is defined to be the \textit{charge conjugation} of  $\Psi$:
\be
\bar{\Psi}:= \Psi^T {\cal C}\,.
\label{11Dcc}
\ee
This is due to the fact that  the three-dimensional Euclidean space does not admit real spinors \textit{i.e.~}Majorana condition.  Here $\cC$ is the  charge conjugation matrix in eleven dimensions  satisfying
\be
\ba{ll}
\cC\Gamma^{M}\cC^{-1}=-(\Gamma^{M})^{T}\,,~~~~&~~~~\cC^{T}=-\cC\,,
\ea
\label{cCcon}
\ee
where $M$ is the eleven-dimensional vector index which decomposes into $\mu=1,2,3$ and  $I=4,5,\cdots,11$.
Throughout the paper,  the complex conjugation of spinors will never appear as we focus on  the Euclidean space.
Further the dynamical spinor field  $\Psi$ has a definite chirality over $(1,2,3)$-space:
\be
\Gamma^{123}\Psi=+i\Psi\,.
\label{chiralPsi}
\ee
In our convention, the field strength is defined by
\be
\tilde{F}_{\mu\nu}{}^{a}{}_{b}=
\partial_{\mu}\tilde{A}_{\nu}{}^{a}{}_{b}-\partial_{\nu}\tilde{A}_{\mu}{}^{a}{}_{b}+
\tilde{A}_{\mu}{}^{a}{}_{c}\tilde{A}_{\nu}{}^{c}{}_{b}-
\tilde{A}_{\nu}{}^{a}{}_{c}\tilde{A}_{\mu}{}^{c}{}_{b}\,,
\label{tF}
\ee
of which the overall sign is opposite to the original convention by Bagger and Lambert \cite{BL}  but faithful to the standard convention.\\

Last but not least, the  Euclidean action (\ref{EuclideanL})  is invariant under the following sixteen  supersymmetry transformation:\footnote{In addition to the \textit{ordinary} supersymmetry (\ref{susyord}), the Euclidean action  enjoys  sixteen  \textit{conformal} supersymmetry~\cite{Bandres:2008vf}, which can be also  twisted  to  define  a novel topological theory on an arbitrary  three-dimensional cone, as was done for $\cN=4$ super Yang-Mills defined on a four-dimensional cone~\cite{Park:2006kt}.}
\be
\ba{l}
\d X^I_a = i\bar{\cE} \G^I \Psi_a \,,
\\
\d \Psi_a = D_\m X_a^I \G^\m \G_I \cE - \frac{1}{6} X^I_b X^J_c X^K_d  f^{bcd}{}_a \G^{IJK} \cE \,,
\\
\d\tilde{A}_{\m ab} = i\bar{\cE} \G_\m \G_I X^I_c \Psi_d f^{cd}{}_{ab} \,,
\ea
\label{susyord}
\ee
which take exactly the same form as in the Minkowskian case. The supersymmetry parameter $\cE$ possesses the opposite chirality compared to  (\ref{chiralPsi}),
\be
\Gamma^{123}\cE=-i\cE\,.
\label{chiralcE}
\ee
\subsection{Description of the twist}
We now come to the description of the twist we perform.  Under $\mbox{Spin}(11)\rightarrow\mbox{Spin(3)}\times\mbox{Spin}(3)\times\mbox{Spin}(5)$,
the eleven-dimensional gamma matrices can be decomposed as
\be
\ba{lll}
\Gamma^{\mu}=\sigma^{\mu}\otimes 1\otimes 1\otimes \sigma^{3}\,,~~~~&~~~~~
\Gamma^{\mu+3}=1\otimes\sigma^{\mu}\otimes 1\otimes\sigma^{1}\,,~~~~&~~~~
\Gamma^{i+6}=1\otimes 1\otimes\gamma^{i}\otimes\sigma^{2}\,,
\ea
\label{forG}
\ee
where $\sigma^{\mu}$, $\mu=1,2,3$  are $2\times 2$ Pauli matrices
\be
\ba{lll}
\sigma^{1}=\left(\ba{cc}0&~1\\1&~0\ea\right)\,,~~~~&~~~~
\sigma^{2}=\left(\ba{cc}0&-i\\+i&0\ea\right)\,,~~~~&~~~~
\sigma^{3}=\left(\ba{cc}+1&\,0\\0&-1\ea\right)\,,
\ea
\ee
and $\gamma^{i}$, $i=1,2,\cdots,5$ are $4\times 4$ gamma matrices in Euclidean five dimensions, satisfying
\be
\ba{ll}
\gamma^{i}\gamma^{j}+\gamma^{j}\gamma^{i}=2\delta^{ij}\,,~~~~&~~~~\gamma^{12345}=1\,.
\ea
\label{fivegamma}
\ee
The charge conjugation matrix in (\ref{cCcon}) takes the explicit  form:
\be
\cC=\epsilon\otimes\epsilon\otimes C\otimes 1\,,
\ee
where $\epsilon=i\sigma^{2}$ as usual, and $C$ is the five-dimensional charge conjugation matrix,
\be
\ba{llll}
\epsilon\sigma^{\mu}\epsilon^{-1}=-(\sigma^{\mu})^{T}\,,~~~~&~~~~C\gamma^{i}C^{-1}=+(\gamma^{i})^{T}\,,
~~~~&~~~~C^{T}=-C\,.
\ea
\ee
The $\so(8)$ chiral matrix is then
\be
\Gamma^{123}=-i\Gamma^{456\cdots 11}=1\otimes1\otimes 1\otimes i\sigma^{3}\,.
\ee
Consequently the eleven-dimensional  spinors carry four indices $\Psi^{\dalpha\dbeta\alpha\pm}$. The first two $\dalpha$, $\dbeta$ indices are for the $\so(3)$ spinor indices and the third one $\alpha$ is for $\so(5)$ spinor indices running from one to four. The last one $\pm$ denotes the $\so(8)$ chirality.  Since the dynamical spinor carries the definite chirality (\ref{chiralPsi}) we have
$~\Psi^{\dalpha\dbeta\alpha-}{=0}$. Similarly from (\ref{chiralcE}), we have  for the supersymmetry parameter  $~\cE^{\dalpha\dbeta\alpha+}{=0}$.
The twist we focus on in the present paper amounts to replacing the three-dimensional rotation group by the diagonal subgroup of $\mbox{Spin(3)}\times\mbox{Spin}(3)$. Accordingly, the twisted spinors admit the following expansion:
\be
\ba{ll}
\Psi^{\da \db \a +} = \textstyle{\frac{1}{\sqrt{2}}} \left( i\eta^\a \e^{\da\db} + \chi_\m^\a (\s^\m \e)^{\da\db}\right)\,,~~~~&~~~~
\cE^{\da \db \a -} = \textstyle{\frac{1}{\sqrt{2}}} \left( i\ve^\a \e^{\da\db}+\ve_\m^\a (\s^\m \e)^{\da\db} \right)\,.
\ea
\label{forSpinor}
\ee
Namely the fermions decompose into a $\SO(3)$ scalar $\eta,\,\ve$ and a one-form  $\chi_{\m}{\rm d}x^{\mu},\,\ve_{\m}{\rm d}x^{\mu}$.
In analogue to (\ref{11Dcc}), we also define the charge conjugation of the $\SO(5)$ spinors for convenience:
\be
\ba{ll}
\bar{\eta}=\eta^{T}C\,,~~~~&~~~~\bar{\chi}_{\mu}=\chi^{T}_{\mu}C\,.
\ea
\ee
Finally for bosons,  our twist  prescribes to decompose the eight bosonic fields   into a $\SO(3)$ one-form and five scalars:
\be
\ba{lll}
X^{I}~~&\longrightarrow&~~\left(\,X_{\mu}{\rm d}x^{\mu}\,,\,Y^{i}\,\right)\,.
\ea
\label{forBoson}
\ee

\subsection{Twisted Lagrangian}
Taking  the decompositions  (\ref{forG}), (\ref{forSpinor}), (\ref{forBoson}) and  an identity (\ref{iden1}) into account,  it is straightforward to rewrite the Euclidean Bagger-Lambert-Gustavsson action (\ref{EuclideanL}) in terms of the anti-commuting fields $\eta,\chi_{\mu}$ and the bosons $A_{\mu},X_{\mu},Y^{i}$. The resulting action defines  our twisted Bagger-Lambert-Gustavsson theory  in three-dimensions:
\be
\ba{ll}
\CS_{\rm twisted}=\displaystyle{\int}{\rm d}^{3}x~\cL_{\rm twisted}\,,~~~~&~~~~
\cL_{\rm twisted}=\cL_{\rm top}+\sqrt{g}L_{g}\,,
\ea
\label{twistedS}
\ee
where
\be
\cL_{\rm top}=
i\epsilon^{\mu\nu\lambda}\! \left(\half A^{+}_{\m ab} \p_\n \tilde{A}^{+ab}_{\l}
+\textstyle{\frac{1}{3}} A^+_{\m ab} \tilde{A}^{+}_{\n}{}^{a}{}_{c}\tilde{A}^{+}_{\l}{}^{cb} \right)
-\epsilon^{\mu\nu\lambda}\Tr\Big[
\half \barchi_{\mu} D_{\nu}^{+} \chi_{\lambda}
-i\half \barchi_{\mu}\gamma^{i} \left[\chi_{\nu},X_{\lambda},Y_{i}\right]
\Big]
\label{Ltoponshell}
\ee
and
\be
\ba{l}
L_{g}= \Tr\Big[~{1\over4}(D_{\mu}X_{\nu}\!-\!D_{\nu}X_{\mu})(D^{\mu}X^{\nu}\!-\!D^{\nu}X^{\mu})
+\half\!
\left(\!D_{\mu}X^{\mu}\!
+i{1\over\,6\sqrt{g}\,}\epsilon^{\mu\nu\lambda}[X_{\mu},X_{\nu},X_{\lambda}]\!\right)^2\\
{}~~~~~~~~~~~~~~~~~~+\half D^{+}_{\mu} Y^{i} D^{-\mu} Y_{i}
+\frac{1}{12}[Y^i, Y^j, Y^k][Y_{i}, Y_{j}, Y_{k}]
+ \frac{1}{4} [X_{\mu} , Y^{j}, Y^{k}][X^{\mu} , Y_{j}, Y_{k}]\\
{}~~~~~~~~~~~~~~~~~~+\bareta D^{-}_{\mu} \chi^{\mu}
+i\bareta\gamma^{i}\left[Y_{i},X_{\mu},\chi^{\mu}\right]
+i\frac{1}{4}\bareta\gamma^{ij}\left[Y_{i},Y_{j},\eta\right]
+i\frac{1}{4}\barchi^{\mu}\gamma^{ij}\left[Y_{i},Y_{j},\chi_{\mu}\right]\Big]\,.
\ea
\label{Ltoponshell2}
\ee

In the above, we have coupled the action to a generic three-dimensional  metric $g_{\mu\nu}$, such that all the derivatives are covariant with respect to both \textit{diffeomorphisms} and \textit{gauge transformations}, and  that $\epsilon^{\lambda\mu\nu}$ is the totally antisymmetric tensor density, satisfying $\e^{123}=1$ and  $\e_{123}=g := \det(g_{\m\n})$.
It is worthwhile to note that  $\cL_{\rm top}$ is manifestly metric-independent as  the Christoffel connection is torsion-free, and that $D_{\mu}X^{\mu}$ is effectively the only term in $L_{g}$ which contains  the Christoffel connection after replacing the fermionic term  $\bareta D^{-}_{\mu} \chi^{\mu}$ by $-\bar{\chi}^{\mu}D^{-}_{\mu}\eta$. The introduction of the curved background metric is necessary for the twisted action to lead to a `topological' theory, in the sense of the metric independence.

Moreover, we have complexified  the gauge field:
\be
\ba{ll}
\tilde{A}^{+}_{\mu ab} :=
\tilde{A}_{\mu ab} -i\textstyle{\frac{1}{\,2\sqrt{g}\,}\,}
\e_{\mu\nu\lambda}X^{\nu}_{c}X^{\lambda}_{d} f^{cd}{}_{ab}\,,~~~~&~~~~
\tilde{A}^{-}_{\mu ab} := \tilde{A}_{\mu ab}
+i\textstyle{\frac{1}{\,2\sqrt{g}\,}\,}\e_{\m\n\l}X^{\n}_{c}X^{\l}_{d} f^{cd}{}_{ab}\,,
\ea
\label{forAmu}
\ee
such that
\be
\ba{ll}
D_{\mu}^{+}=D_{\mu}+i\textstyle{\frac{1}{\,2\sqrt{g}\,}\,}\e_{\m\n\l}\left[X^{\nu},X^{\lambda},{~~~~~}\right]\,,
~~~~&~~~~
D_{\mu}^{-}=D_{\mu}-i\textstyle{\frac{1}{\,2\sqrt{g}\,}\,}
\e_{\m\n\l}\left[X^{\nu},X^{\lambda},{~~~~~}\right]\,,
\ea
\ee
and
\be
\ba{ll}
\tilde{F}^{+}_{\mu\nu}{}^{a}{}_{b} &= \tilde{F}_{\mu\nu}{}^{a}{}_{b}
-i\textstyle{{1\over\sqrt{g}}}\e_{\nu\rho\sigma}(D_{\mu}X^{\rho})_{c}X^{\sigma}_{d}f^{cda}{}_{b}
+i\textstyle{{1\over\sqrt{g}}}\e_{\mu\rho\sigma}(D^{+}_{\nu}X^{\rho})_{c}X^{\sigma}_{d}f^{cda}{}_{b}\\
{}&= \tilde{F}_{\mu\nu}{}^{a}{}_{b}
-i\textstyle{{1\over\sqrt{g}}}\e_{\nu\rho\sigma}(D^{+}_{\mu}X^{\rho})_{c}X^{\sigma}_{d}f^{cda}{}_{b}
+i\textstyle{{1\over\sqrt{g}}}\e_{\mu\rho\sigma}(D_{\nu}X^{\rho})_{c}X^{\sigma}_{d}f^{cda}{}_{b}\,.
\ea
\ee
It is worth while to note:
\be
\ba{l}
D_{\lambda}X_{\mu}-D_{\mu}X_{\lambda}=
D^{+}_{\lambda}X_{\mu}-D^{+}_{\mu}X_{\lambda}=
D^{-}_{\lambda}X_{\mu}-D^{-}_{\mu}X_{\lambda}\,,\\
D^{\mu}X_{\mu}+\textstyle{i{1\over6 \sqrt{g}}}\e^{\m\n\rho}[X_{\m},X_{\n},X_{\rho}]=
D^{+}_{\mu} X^{\mu} -i{1\over\, 3\sqrt{g}\,}\e_{\mu\nu\lambda}[X^\mu, X^\nu, X^\lambda]\,.
\ea
\label{onshellworthnote}
\ee

The Euler-Lagrange equations of motion are, for bosons $X_{\mu},Y^{i},A^{+}_{\mu}$:
\be
\ba{l}
D_{\mu}\left(D^{\lambda}X^{\mu}-D^{\mu}X^{\lambda}\right)
+D^{-\lambda}\left(D^{\mu}X_{\mu}+\textstyle{i{1\over6 \sqrt{g}}}\e^{\m\n\rho}[X_{\m},X_{\n},X_{\rho}]\right)
-i[\bar{\eta},\gamma^{i}\chi^{\lambda},Y_{i}]
\\
+i{1\over\sqrt{g}}\epsilon^{\lambda\mu\nu}\Big(
2[\bar{\eta},X_{\mu},\chi_{\nu}]
-[D^{+}_{\mu}Y^{i},Y_{i},X_{\nu}]
+{1\over2}[\bar{\chi_{\mu}},\gamma^{i}\chi_{\nu},Y_{i}]
\Big)-{1\over2}[Y^{i},Y^{j},[X^{\lambda},Y_{i},Y_{j}]]\!= 0\,,\\
\\
D_{\mu}D^{+\mu}Y^{i}
-i{1\over\sqrt{g}}\epsilon^{\lambda\mu\nu}\Big([D_{\mu}X_{\nu},X_{\lambda},Y_{i}]
-{1\over2}[\bar{\chi_{\mu}},\gamma^{i}\chi_{\nu},X_{\lambda}]\Big)
+i[\bar{\eta},X_{\mu},\gamma^{i}\chi^{\mu}]
\\
-i{1\over2}[\bar{\eta},\gamma^{ij}\eta,Y_{j}]
-i{1\over2}[\bar{\chi}^{\mu},\gamma^{ij}\chi_{\mu},Y_{j}]
+{1\over2}\!\left[Y_{j},Y_{k},[Y^{j},Y^{k},Y^{i}]\right]
+\left[X_{\mu},Y_{j},[X^{\mu},Y^{j},Y^{i}]\right]=0\,,\\
\\
f^{abcd}\Big({1\over2\sqrt{g}}\epsilon^{\mu\nu\lambda}\bar{\chi}_{\nu c}\chi_{\lambda d}
-Y^{i}_{c}D^{+}_{\mu}Y_{i d}
+i{1\over2\sqrt{g}}Y^{i}_{c}\epsilon^{\mu\nu\lambda}[\chi_{\nu},X_{\lambda},Y_{i}]_{d}
-X_{\nu c}(D^{\mu}X^{\nu}-D^{\nu}X^{\mu})_{d}\\
{}~~~~~~~~~~~~
+(D_{\lambda}X^{\lambda}+i{1\over6\sqrt{g}}
\epsilon^{\lambda\mu\nu}[X_{\mu},X_{\nu},X_{\lambda}])_{c}X^{\mu}_{d}
+\bar{\eta}_{c}\chi^{\mu}_{d}\Big)+i\invsg\epsilon^{\mu\nu\lambda} \tilde{F}^{+ab}_{\nu\lambda} = 0\,,
\ea
\label{EOMb}
\ee
and for fermions $\eta,\chi_{\mu}\,$:
\be
\ba{l}
\zeta:=D^{-}_{\m} \chi^{\mu} +i[Y^{i},X^{\mu},\gamma_{i}\chi_{\m}]
+ i{1\over2}[Y^i,Y^j,\gamma_{ij}\eta]=0\,,\\
\xi_{\mu}:= D^{-}_{\m}\eta +i[Y^i,X_\mu,\gamma_i\eta]-i{1\over2}[Y^i,Y^j,\gamma_{ij}\chi_{\mu}]
+\invsg\e_{\m}{}^{\n\l}D^{+}_{\n}\chi_{\l}
-i\invsg\e_{\m\n\l}[Y^i,X^\n,\gamma_{i}\chi^{\l}]=0\,.
\ea
\label{EOMf}
\ee

\subsection{On-shell scalar supersymmetry and BPS equations}
In flat background, the twisted Bagger-Lambert-Gustavsson action~(\ref{twistedS}) is invariant under the sixteen supersymmetries $Q^{\alpha},Q_{\mu}^{\alpha}$  as in the untwisted case. However, in curved backgrounds, in order to have supersymmetry  unbroken,  it is necessary that the corresponding supersymmetry parameters $\ve^{\alpha}$, $\ve_{\mu}^{\alpha}$ should be covariantly constant. Generically, this requirement can be only met for the scalar parameters $\ve^{\alpha}$. Indeed, for our twisted  Bagger-Lambert-Gustavsson theory in a generic curved background,  the twelve vectorial supersymmetries are broken and only the four scalar supersymmetries survive. Explicitly the unbroken scalar supersymmetries are given by:
\be
\ba{cl}
\d X_{\m} \!\!\!\!&= \barve\chi_{\mu}\,,\\

\d Y^i \!\!\!\!&= \barve\gamma^{i}\eta \,, \\

\d \eta \!\!\!\!&= -\left(D_\mu X^{\mu} +i{1\over\, 6\sqrt{g}\,}\e_{\mu\nu\lambda}[X^\mu, X^\nu, X^\lambda]\right)\ve
+i{1\over6}[Y^i, Y^j, Y^k]\gamma_{ijk}\ve \,,\\

\d \chi_{\lambda} \!\!\!\!&= \invsg\e_{\lambda\mu\nu} D^{\mu} X^{\nu} \ve +
D^{+}_{\lambda} Y^{i}\gamma_{i}\ve
+i{1\over2}[Y^i, Y^j, X_{\lambda}]\gamma_{ij}\ve\,,\\

\d \tilde{A}_{\m ab} \!\!\!\!&= i\left(-X_{\mu c} \barve\eta_{d}
+\invsg\epsilon_{\mu\nu\lambda} X^{\nu}_{c} \barve\chi^{\lambda}_{d}
+Y_{i c} \barve\gamma^{i}\chi_{\m d}\right)f^{cd}{}_{ab} \; .
\ea
\label{onSUSY}
\ee
Equivalently in terms of  scalar supercharges:
\be
\label{onSUSY2}
\ba{cl}
\left[Q^\a, X_{\m}\right] \!\!\!&=  \chi_{\m}^\a \,,\\
\left[Q^\a, Y^i \right] \!\!\!&= (\gamma^{i}\eta)^{\alpha}\,,\\
\{ Q^\a, \bareta_{\beta} \} \!\!\!&= -\left(D_\mu X^{\mu} +i{1\over\, 6\sqrt{g}\,}\e_{\mu\nu\lambda}[X^\mu, X^\nu, X^\lambda]\right)\delta^\a{}_\b
-i{1\over6}[Y^i,Y^j,Y^k](\gamma_{ijk})^{\a}{}_{\b}\,,\\
\{ Q^\a, \barchi_{\l\b} \} \!\!\!&= \invsg\e_{\l}{}^{\m\n} D_\m X_\n \delta^{\a}{}_{\b} +
D^{+}_\l Y^i(\gamma_i)^{\a}{}_{\b}
-i{1\over2}[Y^i,Y^j,X_\l](\gamma_{ij})^{\a}{}_{\b}\,,\\
\left[ Q^\a, \tilde{A}_{\m ab}\right]\!\!\!&=
i\Big(-X_{\m c} \eta^\a_d+\invsg\e_{\m}{}^{\n\l} X_{\n c}  \chi^\a_{\l d} +  Y_{ic} (\gamma^{i}\chi_{\m d})^{\alpha}
\Big)f^{cd}{}_{ab}\;.
\ea
\label{onshellQ}
\ee
Successive scalar supersymmetry transformations give
\be
\ba{lll}
\left[\{Q^\a,Q^\b\},X_\m\right] &=& i[Y^i,Y^j,X_\m](\gamma_{ij}C^{-1})^{\a\b}\,,
\\
\left[\{Q^\a,Q^\b\},Y^i\right] &=& i[Y^i,Y^j,Y^i](\gamma_{ij}C^{-1})^{\a\b}\,,
\\
\left[\{Q^\a,Q^\b\},\eta_\gamma\right] &=&  i[Y^i,Y^j,\eta_{\gamma}](\gamma_{ij}C^{-1})^{\a\b}
+\delta^\beta{}_\gamma \zeta^{\alpha} + \delta^\alpha{}_\gamma \zeta^{\beta}\,,
\\
\left[\{Q^\a,Q^\b\},\chi_{\m\g}\right] &=& i[Y^i,Y^j,\chi_{\m\gamma}](\gamma_{ij}C^{-1})^{\a\b}
-\delta^{\beta}{}_{\gamma} \xi^{\alpha}_{\mu} - \delta^{\alpha}{}_{\gamma} \xi^{\beta}_{\mu} \,,
\\
\left[\{Q^\a,Q^\b\},\tilde{A}_{\m a b}\right] &=& 2i \left( Y^{i}_{c}D_{\m} Y^{j}_{d}(\gamma_{ij}C^{-1})^{\alpha\beta}\right) f^{cd}{}_{ab} \,.
\ea
\label{QQon}
\ee
Apart from the Euler-Lagrange equations of the fermions, $\zeta,\xi_{\mu}$ (\ref{EOMf}),
the right hand sides in (\ref{QQon}) correspond precisely to the gauge transformation (\ref{gaugetr}). Thus, \textit{the scalar supercharges are nilpotent on-shell up to gauge transformations.}\\

From the supersymmetry transformations of the fermions (\ref{onSUSY}), we see that  supersymmetric invariant bosonic configurations must  satisfy the following BPS conditions:
\footnote{For various BPS states in the original untwisted BLG theory, including the classification, we refer \cite{bps1,bps2,Krishnan:2008zm}.}
\be
\ba{ll}
D^{+}_{\mu} X^{\mu} -i{1\over\, 3\sqrt{g}\,}\e_{\mu\nu\lambda}[X^\mu, X^\nu, X^\lambda]=0\,,~~~~&~~~~
D^{+}_{\mu}X_{\nu}-D^{+}_{\nu}X_{\mu}=0\,,
\\
D^{+}_{\mu} Y^{i}=0\,,~~~~~~~~~~~~~~~[Y^i, Y^j, Y^k]=0\,,~~~~&~~~~[Y^i, Y^j, X_{\lambda}]=0\,.
\ea
\ee
Further, these BPS conditions imply  the bosonic Euler-Lagrange equations of motion (\ref{EOMb}) if and only if
\be
F^{+}_{\mu\nu}=0\,.
\ee

\section{Off-shell supersymmetric formulation of the twist\label{secoffshell}}
The above on-shell formulation of the  twist is not yet sufficient to define  a genuine topological field theory which depends  only on the topology of  the three-dimensional base manifold,  since the scalar supercharges are only on-shell nilpotent and the scalar supersymmetry transformations (\ref{onshellQ}) are not independent from the base manifold metric.  In this section we construct an off-shell supersymmetric formalism  of the  twist which will eventually lead to  a genuine topological field theory.

\subsection{Off-shell supersymmetry algebra}
Our off-shell supersymmetric formulation  requires two auxiliary fields which we call  $h$ and $h_\mu$. The off-shell $Q$-variations are defined over $\{X_{\mu},Y^{i},h,h_{\mu},\eta,\chi_{\mu},A^{+}_{\mu}\}$ as follows:\footnote{Transforming (\ref{onSUSY2}) to (\ref{Q-off}),
we made the identification,
\[
h \equiv D^{+\mu}X_{\mu}-i\textstyle{\frac{1}{\,3\sqrt{g}}} \e^{\m\n\l} [X_\m,X_\n,X_\l]  \,, ~~~~~~~~h_\m \equiv \invsg\e_{\m\n\l}D^{+\nu}X^{\lambda} \,,
\]
where ``$\equiv$'' means on-shell equivalence. To obtain
the $Q$-variation
of the auxiliary fields, we take the variation of their on-shell
values and use the equations of motion to remove any metric-dependent terms.}
\be
\label{Q-off}
\ba{cll}
\left[Q^\a, X_{\m}\right] &=&  \chi_{\m}^\a \,,\\
\left[Q^\a, Y^i \right] &=& (\gamma^{i}\eta)^{\alpha}\,,\\
\left[Q^\a, h \right] &=& -i{1\over2}[Y^{i},Y^{j},(\gamma_{ij}\eta)^{\alpha}]\,,
\\
\left[Q^\a, h_{\mu} \right] &=& -D^{+}_{\mu} \eta^{\alpha} +i[X_{\mu},Y^{i},(\gamma_{i} \eta)^{\alpha}] +i{1\over2}[Y^{i},Y^{j},(\gamma_{ij}\chi_{\mu})^{\alpha}]\,,
\\
\{ Q^\a, \bareta_{\beta} \} &=& -h\delta^{\alpha}_{~\beta} -i{1\over6}[Y^i,Y^j,Y^k](\gamma_{ijk})^{\a}_{~\b}\,,
\\
\{ Q^{\alpha}, \barchi_{\mu\beta} \} &=&   h_\mu \delta^{\a}_{~\b} + D^{+}_{\mu}Y^{i}(\gamma_{i})^{\alpha}{}_{\beta}
 -i{1\over2}[Y^i,Y^j,X_{\mu}](\gamma_{ij})^{\a}_{~\b}\,,\\
\left[ Q^\a, \tilde{A}^{+}_{\m ab}\right] &=& i\left(-X_{\m c} \eta^\a_d
+Y_{ic}(\gamma^{i}\chi_{\mu d})^{\alpha}
\right)f^{cd}{}_{ab} \; .
\ea
\label{offshellQ}
\ee
It is straightforward to verify that our $Q$-variations (\ref{Q-off})
are nilpotent strictly off-shell, up to a gauge transformation:
for all the fields in $\{X_{\mu},Y^{i},h,h_{\mu},\eta,\chi_{\mu},A^{+}_{\mu}\}$, we find
\be
Q^{2}=\mbox{gauge~transformation}\,,
\ee
where, with an arbitrary  constant $c$-number spinor $v_{\alpha}$, $Q = \bar{v}_{\alpha}Q^{\alpha}$ and
the gauge parameter (\ref{gaugetr}) is given by
\be
\L_{ab} = i\half Y^i_a Y^j_b (\g_{ij}C^{-1})^{\a\b} \bar{v}_\a \bar{v}_\b\,.
\ee

Here the off-shell supersymmetry algebra is defined for $\tilde{A}^{+}_{\m ab}$ and   not for $\tilde{A}^{-}_{\m ab}$. In our off-shell supersymmetric formalism it is not necessary to define   the $Q^{\alpha}$-variation of $\tilde{A}^{-}_{\m ab}=(\tilde{A}^{+}_{\m ab})^{\ast}$.  In fact,  in our off-shell supersymmetric formulation  we may relax  the decomposition rule  of the complex gauge field into the real and imaginary parts given in Eq.~(\ref{forAmu}), such that  $\tilde{A}_{\mu ab}$  will never appear and
we may keep only the reality condition $\tilde{A}^{-}_{\m ab}=(\tilde{A}^{+}_{\m ab})^{\ast}$.\footnote{Note that $[ Q^\a, \tilde{A}^{+}_{\m ab}]^{\dagger}$ does not lead to the $Q^{\alpha}$-variation of $\tilde{A}^{-}_{\m ab}=(\tilde{A}^{+}_{\m ab})^{\ast}$, since $Q^{\alpha}$ is not hermitian.} In this case, the identities (\ref{onshellworthnote}) do not hold anymore.

\paragraph{Ghost number}
In topological field theories, it is often useful  to introduce
the so-called ghost-number $U$, though it may {\em not} lead to a symmetry of the topological action, as will be the case with our  twisted Lagrangian. We first assign ghost number one to the scalar supercharges, $U(Q)=1$. Then (\ref{Q-off}) uniquely determines the ghost number of each field:
\be
U(X_\m,\,\chi_\m,\,h_\m,\, Y,\,\eta,\,h,\,\tilde{A}_\m^{+}) = (-1,\,0,\,+1,\,+1,\,+2,\,+3,\,0)\,.
\ee

\subsection{Off-shell supersymmetric Lagrangian}
Provided  the off-shell supersymmetry algebra,
it is straightforward  to obtain the off-shell supersymmetric Lagrangian:
\be
\label{Lag2}
\ba{ll}
\cL_{\rm off-shell} =&\!\!
i\epsilon^{\mu\nu\lambda} \left(\half A^{+}_{\m ab} \p_\n \tilde{A}^{+ab}_{\l}
+\textstyle{\frac{1}{3}} A^+_{\m ab} \tilde{A}^{+}_{\n}{}^{a}{}_{c}\tilde{A}^{+}_{\l}{}^{cb} \right)
\\
&-\epsilon^{\mu\nu\lambda}\Tr\Big(\frac{1}{2} \barchi_{\mu} D_{\nu}^{+} \chi_{\lambda}
-i\frac{1}{2} \barchi_{\mu}\left[\gamma_{i}\chi_{\nu},X_{\lambda},Y^{i}\right]
+i{1\over 3}h[X_{\mu},X_{\nu},X_{\lambda}]
-h_{\mu}D^{+}_{\nu} X_{\lambda}\Big)
\\{}&
+\sqrt{g}\,\Tr\Big(\half D^{+}_{\mu} Y^iD^{-}_{\mu} Y_i
-\half h^2+hD^{+\mu}X_{\mu}
-\half h^{\mu}h_{\mu}
\\
{}&~~~~~~~~~~~~~~~~~+\frac{1}{12}[Y^i, Y^j, Y^k][Y_i, Y_j, Y_k]
+ \frac{1}{4} [X_{\mu}, Y^{j}, Y^{k}][X^{\mu} , Y^{j}, Y^{k}]\\
{}&~~~~~~~~~~~~~~~~~-\bar{\chi}^{\mu}D^{-}_{\mu}\eta
+i\bareta\gamma_{i}\left[Y^{i},X_{\mu},\chi^{\mu}\right]
+i\frac{1}{4}\bareta\gamma_{ij}\left[Y^{i},Y^{j},\eta\right]
+i\frac{1}{4}\barchi^{\mu}\gamma_{ij}\left[Y^{i},Y^{j},\chi_{\mu}\right]\Big)\,.
\ea
\ee
Integrating  out the auxiliary fields, the above off-shell supersymmetric Lagrangian (\ref{Lag2})  reduces to the form:
\be
\ba{ll}
\cL_{\rm off-shell} \equiv&
i\epsilon^{\mu\nu\lambda} \left(\half A^{+}_{\m ab} \p_\n \tilde{A}^{+ab}_{\l}
+\textstyle{\frac{1}{3}} A^+_{\m ab} \tilde{A}^{+}_{\n}{}^{a}{}_{c}\tilde{A}^{+}_{\l}{}^{cb} \right)
-\epsilon^{\mu\nu\lambda}\Tr\Big(\frac{1}{2} \barchi_{\mu} D_{\nu}^{+} \chi_{\lambda}
-i\frac{1}{2} \barchi_{\mu}\left[\gamma_{i}\chi_{\nu},X_{\lambda},Y^{i}\right]\Big)
\\{}&\!\!\!
+\sqrt{g}\,\Tr\Big(
\half(D^{+\mu}X_{\mu}\!-i{1\over3\sqrt{g}}\epsilon^{\mu\nu\lambda}
[X_{\mu},X_{\nu},X_{\lambda}])^2
+{1\over4}(D^{+}_{\mu}X_{\nu}-D^{+}_{\nu}X_{\mu})(D^{+\mu}X^{\nu}-D^{+\nu}X^{\mu})
\\{}&~~~~~~~~~~~~~~~
+\half D^{+}_{\mu} Y^iD^{-}_{\mu} Y_i
+\frac{1}{12}[Y^i, Y^j, Y^k][Y_i, Y_j, Y_k]
+ \frac{1}{4} [X_{\mu}, Y^{j}, Y^{k}][X^{\mu} , Y^{j}, Y^{k}]
\\{}&~~~~~~~~~~~~~~~-\bar{\chi}^{\mu}D^{-}_{\mu}\eta
+i\bareta\gamma_{i}\left[Y^{i},X_{\mu},\chi^{\mu}\right]
+i\frac{1}{4}\bareta\gamma_{ij}\left[Y^{i},Y^{j},\eta\right]
+i\frac{1}{4}\barchi^{\mu}\gamma_{ij}\left[Y^{i},Y^{j},\chi_{\mu}\right]
\Big)\,,
\ea
\ee
which is very similar, but not identical, to the on-shell supersymmetric Lagrangian (\ref{Ltoponshell}), (\ref{Ltoponshell2}). Only if we assume the decomposition of the complex gauge field into the real and imaginary parts  given in (\ref{forAmu}), they coincide.

A crucial feature of the off-shell supersymmetric Lagrangian (\ref{Lag2}) is that it can be written
as a sum of   $Q$-closed and $Q$-exact  parts:
\be
\cL_{\rm off-shell} = \CL_{{\rm closed}}+\{ Q , \S \}\,,
\ee
where, firstly with a pair of arbitrary constant $c$-number spinors $\bar{v}_{\alpha}, u^{\beta}$ satisfying $\bar{v}_{\alpha}u^{\alpha}\neq 0$, the scalar supercharge and
and the fermionic scalar in the $Q$-exact part are
\be
\ba{ll}
Q=\bar{v}_{\alpha}Q^{\alpha}\,,~~~~&~~~~
\S = \bar{\S}_\a u^\a / (\bar{v}_\b u^\b)\,,
\ea
\label{linearcom}
\ee
of which  the fermionic $\SO(5)$ spinor is given by
\be
\bar{\S} = \half h \bareta -\half h^\m \barchi_{\m} +\half (D^{+}_{\m}Y^i)\barchi^{\mu}\g_i
-(D^{+}_{\m}X^\m) \bareta
-i\textstyle{\frac{1}{4}}[Y^i,Y^j,X_\m]\bar{\chi}^\m\g_{ij}
-i\textstyle{\frac{1}{12}} [Y^i,Y^j,Y^k] \bareta\g_{ijk}\,.
\ee
The $Q$-closed part  is then
\be
\label{Ltop1}
\ba{ll}
\cL_{\rm closed} =& i\e^{\mu\nu\lambda}({1\over2}f^{abcd}A^{+}_{\mu ab}\partial_{\nu}A^{+}_{\lambda cd}-{1\over3}f^{cdag}f^{efb}_{~~~~~g}A^{+}_{\mu ab}
A^{+}_{\nu cd}A^{+}_{\lambda ef})
\\&
+\epsilon^{\mu\nu\lambda} \Tr\Big(-{1\over2}\barchi_{\mu}D^{+}_{\nu}\chi_{\lambda}
+ h_{\mu}D^{+}_{\nu}X_{\lambda}
+i{1\over2}\barchi_{\mu}[\gamma_{i}\chi_{\nu},X_{\lambda},Y^{i}]\\
&~~~~~~~~~~~~~~~~~-i{1\over3}h[X_{\mu},X_{\nu},X_{\lambda}]
-i\bareta[\chi_{\mu},X_{\nu},X_{\lambda}]
- {i\over2}D^{+}_{\mu}Y^{i}[X_{\nu},X_{\lambda},Y_{i}]\,\Big)\,.
\ea
\ee
Direct manipulation indeed shows that $\cL_{\rm closed}$ is  $Q$-closed up to total derivative terms, and more interestingly about the $Q$-exact term,
\be
\ba{ll}
\{Q^{\alpha},\bar{\Sigma}_{\beta}\}&= \delta^{\alpha}_{~\beta}\{Q,\Sigma\}\\
{}&=\delta^{\alpha}_{~\beta}\Tr\!\Big[
\half D^{+}_{\mu}Y^{i}D^{+\mu}Y_{i} -\half h^{2} +h D^{+\mu}X_{\mu} -\half h^{\mu}h_{\mu}
-\bar{\chi}^{\mu}D^{-}_{\mu}\eta
+i\bar{\eta}\gamma^{i}[Y^{i},X_{\mu},\chi^{\mu}]
\\&~~~~~~~~~~~~~~~
+i{1\over4}\bar{\eta}\gamma_{ij}[Y^{i},Y^{j},\eta]
+i\invsg\epsilon^{\mu\nu\lambda}\bar{\eta}[\chi_{\mu},X_{\nu},X_{\lambda}]
+i{1\over4}\bar{\chi}^{\mu}\gamma_{ij}[Y^{i},Y^{j},\chi_{\mu}]\\
&~~~~~~~~~~~~~~~+{1\over12}[Y^{i},Y^{j},Y^{k}][Y_{i},Y_{j},Y_{k}]
+{1\over4}[Y^{i},Y^{j},X^{\mu}][Y_{i},Y_{j},X_{\mu}]\Big]\,.
\ea
\label{QSd}
\ee
In fact,  utilizing the existing $\SO(5)$ symmetry of the action, one can rotate the
constant $c$-number spinor such that only one component is nontrivial \textit{e.g.~}$\bar{v}_{\alpha}=v\delta_{\alpha 1}$. In this case, from (\ref{QSd}) only the corresponding one component of $\bar{\S}_{\alpha}$,
\textit{i.e.~}$\bar{\Sigma}_{1}$ couples to the supercharge and contributes to the formation of the $Q$-exact part of the Lagrangian. In this way, different choices of the linear combination of the four scalar supercharges (\ref{linearcom}) are all $\SO(5)$ equivalent.  
At this point it is worthwhile to compare with a  topological twisting of $\cN=4$ super Yang-Mills theory \cite{Marcus:1995mq,Kapustin:2006pk,Park:2006kt} where there appears a pair of scalar supercharges.  In contrast to our case, the twisted action possesses no $R$-symmetry which would rotate  the two supercharges to each other. Hence, a different linear combination of the scalar supercharges defines inequivalent   cohomology.

The $Q$-closed part  has the ghost number zero and  contains no metric dependent term, being explicitly topological, \textit{c.f.~}(\ref{Ltoponshell}).
On the other hand, the $Q$-exact part has no definite ghost number and contains explicitly metric dependent terms. In fact, all the metric dependence of the off-shell supersymmetric Lagrangian (\ref{Lag2}) can be read off from $\bar{\Sigma}$, because the $Q$-transformations (\ref{offshellQ}) are independent of the base manifold metric. Thus, the energy-momentum tensor is $Q$-exact, and our off-shell supersymmetric formulation of the Bagger-Lambert-Gustavsson action indeed defines a genuine topological field theory in three dimensions.  Note that since $\bar{\Sigma}$ does not involve $A^{-}_{\mu}$, our  $Q$-transformation rule - which is defined over
$\{X_{\mu},Y^{i},h,h_{\mu},\eta,\chi_{\mu},A^{+}_{\mu}\}$ only -  can be applied to it.

Fierz identities have been heavily used for the derivation of the above formulae.
We summarize them in  the appendix.

\section{Observables and partition function\label{secobs} }

\subsection{Observables}

As is well-known, a local operator
that is $Q$-closed up to  total derivatives
leads to a series of observables.
For instance, we can have a relation:
\be
\ba{llll}
\left[Q, \CO_3 \right] = {\rm{d}} \CO_2 \,,~~~~&~~~~
\{ Q, \CO_2 \} = {\rm{d}} \CO_1 \,,~~~~&~~~~
\left[ Q, \CO_1 \right] = {\rm{d}} \CO_0 \,,~~~~&~~~~
\{ Q, \CO_0 \} = 0 \,.
\ea
\ee
Here, $\CO_n$ are $n$-forms with alternating statistics.
The first relation holds by assumption, and the rest follows
from the nilpotency of $Q$.
The integration of $\CO_n$   over an $n$-cycle then
gives a well-defined observable.

One particular family of observables that comes free
for any topological theory is the one associated with the $Q$-closed part of the off-shell supersymmetric Lagrangian. For our theory, we find
\be
\ba{lll}
\left[Q^{\alpha},\cL_{\rm closed}\right] = \partial_{\mu}\cL^{\alpha\mu}\,,~~~~~&~~~~~
\left[Q^{(\alpha},\cL^{\beta)\mu}\right] = \partial_{\nu}\cL^{\alpha\beta\mu\nu}\,,
~~~~~&~~~~~
\left[Q^{(\alpha},\cL^{\beta\gamma)}_{\mu\nu}\right]=0\,,
\ea
\ee
where the brackets denote the symmetrization of the spinorial indices with weight one,
\textit{i.e.~}$A^{(\alpha} B^{\beta)}=\half (A^\a B^\b + A^\b B^\a)$  and
\be
\ba{ll}
\cL^{\alpha}_{\mu}\!\!\! &= -\epsilon_{\mu\nu\lambda}\Tr\Big(
i{1\over4}[Y^{i},Y^{j},(\gamma_{ij}\chi^{\nu})^{\alpha}]X^{\lambda}
+i{1\over2}[X^{\nu},Y^{i},(\gamma_{i}\eta)^{\alpha}]X^{\lambda}
+{1\over2}\eta^{\alpha}D^{+\nu}X^{\lambda}
-{1\over2}\chi^{\alpha\nu}h^{\lambda}\Big)\, ,
\\
\cL^{\alpha\beta\mu\nu}\!\!\! &= \epsilon^{\mu\nu\lambda}
\Tr\Big({1\over2}\eta^{(\alpha}\chi^{\beta)}_{\lambda}
+i{1\over12}[Y^{i},Y^{j},Y^{k}]X_{\lambda}(\gamma_{ijk}C)^{\alpha\beta}\Big)\,.
\ea
\ee

\paragraph{Wilson loop}

Wilson loop operator is one of the most fundamental observables in
any non-Abelian gauge theory. Moreover, the Wilson loop in pure Chern-Simons
theory \cite{purecs} has been used to compute knot invariants
of three manifolds.
So, it is natural to ask whether
we can have a sort of Wilson loop as an observable in our case too.

The simplest Wilson loop made of $\tilde{A}^+$ is not a good observable
since $\tilde{A}^+$ is not $Q$-closed,
\[
\left[ Q^\a, \tilde{A}^{+}_{\m ab}\right] = i\left(-X_{\m c} \eta^\a_d
+Y_{ic}(\gamma^{i}\chi_{\mu d})^{\alpha}
\right)f^{cd}{}_{ab} \,.
\]
It is tempting to modify $\tilde{A}^+$ further to make it $Q$-closed,
but it appears that it is not possible to do so and there is no Wilson loop-like observable in the twisted Bagger-Lambert-Gustavsson theory.
The argument goes as follows.
Consider
$\tilde{A}_{\m ab}^+~\goto~
\CA_{\m ab} = \tilde{A}_{\m ab}^+ + B_{\m ab}$ with
$B_{\m ab} \equiv i f^{cd}{}_{ab} X_{\m c} Y_{id} s^i$,
where $s^i$ is a constant vector of $\mbox{SO}(5)$.
Requiring closedness of $\CA_{\m ab}$ under $v_\a Q^\a$, we find two conditions:
\[{\rm (a)}~\;\;\; s^i v_\a = v_\b (\g^i)^\b{}_\a\,,
\;\;\;\;\;~~~
{\rm (b)}~\;\;\; v_\a = s^i v_\b (\g^i)^\b{}_\a\,.
\]
Condition (b) is not very strong; for a given $v_\a$,
it is easy to choose an $s^i$ satisfying it.
On the other hand, condition (a) is very strong.
Using the $\SO(5)$ covariance, we can always go to a basis in which,
say, $s^1=0$. Then, we have
$0 = v_\b (\g^1)^\b{}_\a$,
which implies an unacceptable condition  $v_\a=0$ because $\g^1$ is invertible.

\subsection{Partition function\label{secpar}}
Let us now consider the partition function in a quantum field theory in general:
\be
\label{ZZ}
{\cal Z} = \int \cD\Phi~\exp(- \cS) \, .
\ee
In the usual semi-classical expansion, one proceeds in four steps: classical action,
one loop determinants,  higher loop corrections and non-perturbative instanton corrections. However,
in topological quantum field theory, one loop correction alone around all the BPS configurations can lead to an exact result due to the localization.

The localization follows from the fact that
the partition function (\ref{ZZ}) and the vev of observables
are invariant under the smooth deformation of the Lagrangian,
\be
\cL = \CL_{{\rm closed}}+\{ Q , \S \}
\;\;\; \rightarrow \;\;\;
\cL_t = \CL_{{\rm closed}}+ t \{ Q , \S \}\,,
\ee
with an arbitrary real parameter $t$. For large positive values
of $t$, the path integral is ``localized'' to field configurations
with $\{Q, \Sigma\}=0$. The expression for $\{Q, \Sigma\}$ (\ref{QSd}) implies that the path integral localizes to
\be
\ba{lllll}
D^{+}_{\mu} X^{\mu} =0\,,~~~&~~~
h_{\mu} =0\,,~~~&~~~
D^{+}_{\mu} Y^{i}=0\,,~~~&~~~[Y^i, Y^j, Y^k]=0\,,~~~&~~~[Y^i, Y^j, X_{\lambda}]=0\,.
\ea
\ee

A systematic study of the full partition function,
including the integral over the BPS configurations, is out of the scope of this work.
Even at the classical level, we would have to deal with the subtleties
due to the imaginary value and gauge non-invariance
of the Chern-Simons term; see \cite{M-inst} for a recent discussion.
Here, as a first step forward,
we evaluate the one loop determinants
in the trivial background with vanishing vev for all fields.

When the three-algebra is equipped with a positive definite norm, the possible dimension of the three-algebra is either one (trivial case), four or infinity. Either precisely for the trivial dimension, or effectively for the evaluation of the one-loop determinants in the nontrivial dimensions, we have  copies
of the following action for free fields (in the form notation),
\be
\label{1-loop:S}
\displaystyle{\cS=\!\int\!{\rm d}^{3}x}\,
\displaystyle{{\sqrt{g}\Big[\half \left(X_{1},\Delta_{1}X_{1}\right)
+\half(Y^{i},\Delta_{0}Y_{i})
-(\bar{\chi}_{1},\ast{\rm d}\chi_{1})
-(\bar{\eta},{\rm d}^{\dagger}\chi_{1})
\Big]\,,}}
\ee
where $\Delta_{0}$ and $\Delta_{1}$ are Laplacians acting on zero and one forms respectively,
\be
\ba{ll}
\Delta_{0}Y^{i}=-\nabla^{\mu}\nabla_{\mu}Y^{i}\,,~~~~&~~~~\Delta_{1}X_{\mu}
=-\nabla^{\nu}\nabla_{\nu}X_{\mu}-[\nabla_{\mu},\nabla^{\nu}]X_{\nu}\,.
\ea
\ee
Note also that $\sqrt{g}\,(\bar{\chi}_{1},\ast{\rm d}\chi_{1})=\bar{\chi}_{1}\wedge{\rm d}\chi_{1}$.
For the nontrivial  three-algebra dimensions
we omitted the Chern-Simons term of the gauge field, since
at one loop level the contribution from the gauge field
cancels out against those from the gauge-fixing ghosts.

In general, according to the Hodge  theorem, any $p$-form, $\psi_{p}$, in a compact  manifold of the positive definite signature decomposes uniquely into the harmonic form, $h_{p}$, exact form, ${\rm d}\alpha_{{{p-1}}}$, and coexact form, ${\rm d}^{\dagger}
\beta_{{{p+1}}}=(-1)^{p+1}\ast{\rm d}\ast\beta_{p+1}$,
\be
\psi_{p}=h_{p}+{\rm d}\alpha_{{{p-1}}}+{\rm d}^{\dagger}
\beta_{{{p+1}}}\,,
\ee
where $h_{p}$, $\alpha_{p-1}$ and $\beta_{p+1}$ are all globally well defined. From the positive definiteness, we also have ${\rm d}h_{p}=0$, ${\rm d}^{\dagger}h_{p}=0$. The Laplacian on $p$-form \textit{i.e.~}$\Delta_{p}$ is given by\footnote{
Explicitly we have for a $p$-form, $\psi$,
\[
\ba{ll}
({\rm d}\psi)_{a_{1}a_{2}\cdots a_{p+1}}=(p+1)\nabla_{[a_{1}}\psi_{a_{2}a_{3}\cdots a_{p+1}]}\,,~~~&~~~
({\rm d}^{\dagger}\psi)_{a_{1}a_{2}\cdots a_{p-1}}=-\nabla^{b}\psi_{b\,a_{1}a_{2}\cdots a_{p-1}}\,,\\
\multicolumn{2}{l}{
\displaystyle{(\Delta_{p}\psi)_{a_{1}a_{2}\cdots a_{p}}=
-\nabla^{b}\nabla_{b}\psi_{a_{1}a_{2}\cdots a_{p}}+p[\nabla_{b},\nabla_{[a_{1}}]
\psi^{b}{}_{a_{2}a_{3}\cdots a_{p}]}}\,.}
\ea
\]
}
\be
\Delta_{p}={\rm d}^{\dagger}{\rm d}+{\rm d}{\rm d}^{\dagger}\,,
\ee
so that each of ${\rm d}^{\dagger}{\rm d}$ and ${\rm d}{\rm d}^{\dagger}$ diagonalizes  over  the harmonic, exact and coexact $p$-form spaces.

In our case of the free action above (\ref{1-loop:S}), integrating out ${\eta}$ field forces to set ${\rm d}^{\dagger}\chi_{1}=0$, and hence with $\chi_{1}=h_{1}+{\rm d}\alpha_{{{0}}}+{\rm d}^{\dagger}\beta_{{{2}}}$, from the positive definiteness  the partition function saturates at
\be
\ba{lll}
{\rm d}\alpha_{0}=0~~~&\mbox{for}&~~~\chi_{1}\,,
\ea
\ee
which can be regarded as   a gauge fixing in   BRST quantization.

When there are fermionic zero modes,
the bare partition function vanishes.
In our case, there are one zero mode for $\eta$ and $b_1$ (the first Betti number) zero modes for $\chi_1$.
Assuming that the right number of zero modes are absorbed by
products of fermions from observables and/or interaction vertices,
we find
\be
\displaystyle{{\cal Z}_{\rm one-loop}:=\int\!{\cD X\cD Y\cD \chi}~e^{-\cS}=
\frac{\mbox{Pf}\,[C(\ast{\rm d})_{1}]}{\,\left[\det\Delta_{0}\right]^{\frac52}\left[\det\Delta_{1}\right]^{\frac12}\,}
=\frac{[\det\Delta_{1}]^{\frac{3}{2}}}{[\det\Delta_{0}]^{\frac{9}{2}}\,}\,,}
\ee
where the second equality follows from
\footnote{
In general for a $p$-form in $d$ dimension, we have
\[
\ba{lll}
\det\Delta_{p}=\det\Delta_{d-p}\,,~~~~&~~~~
\det\Delta_{p}=\det({\rm d}^{\dagger}{\rm d})_{p}\det({\rm d}{\rm d}^{\dagger})_{p}\,,~~~~&~~~~\det({\rm d}^{\dagger}{\rm d})_{p}=\det({\rm d}{\rm d}^{\dagger})_{p+1}\,.
\ea
\]
}
$\mbox{Pf}=\sqrt{\det}$, $C^{2}=-1_{{\scriptscriptstyle{4\times 4}}}$, $\det({\rm d}^{\dagger}{\rm d})_{1}=\det\Delta_{1}/\det\Delta_{0}$. The final expression  is nothing but the topological quantity known as the Ray-Singer torsion in three-dimensions:
\be
\displaystyle{
{\prod_{p=0}^{3}\,\left[\det\Delta_{p}\right]^{-(-1)^{p} \frac{1}{2}p}\,}=
\frac{\left[\det\Delta_{0}\right]^{\frac{3}{2}}}{\left[\det\Delta_{1}\right]^{1\over2}}=
\cZ_{\rm one-loop}^{\,-\frac{1}{3}}}\,.
\ee

We close this section with a comparison with
a similar computation in the Rozansky-Witten theory~\cite{roz}.
The fermionic part of our free action (\ref{1-loop:S})
is essentially identical to that of Rozansky-Witten theory.
The bosonic part of Rozansky-Witten theory
is a non-linear sigma model with a hyper-K\"ahler target space,
so it is quite different from our theory.
Nevertheless, the combination
of the bosonic and fermionic contributions
of Rozansky-Witten theory also gives rise to the Ray-Singer torsion but with a different power from ours, \textit{i.e.~}$-1/2$ versus $-1/3$.

\section{Relation to M5: partial topological twist of six-dimensional theory\label{secM5}}
If we introduce an auxiliary three manifold,  an explicit realization of an infinite dimensional   three-algebra follows straightforwardly from the Nambu three-bracket defined on the  internal manifold. This suggests that Bagger-Lambert-Gustavsson theory with infinite dimensional gauge group  describes M5-brane as a condensation of multiple M2-branes~\cite{Ho:2008nn,Park:2008qe,Bandos:2008fr}. In fact,  by generalizing the Brink-Di Vecchia-Howe-Polyakov method, Nambu-Goto action for a five-brane can be reformulated as a
three-dimensional gauged nonlinear sigma model  having a Nambu three-bracket squared potential~\cite{Park:2008qe}.

Introducing  a functional basis for the three-manifold $T^{a}(y)$, we let all the variables be functions on the whole six-dimensions \textit{e.g.~}$X_{\mu}(x,y)=X_{\mu a}(x)T^{a}(y)$. We represent the three-algebra by
\be
\ba{cll}
[X,Y,Z]~&\equiv&~\textstyle{\frac{1}{\sqrt{\hat{g}}}}\epsilon^{\hl\hm\hn}
\partial_{\hl}X\partial_{\hm}Y\partial_{\hn}Z\,,\\
D_{\mu}X~&\equiv&~\partial_{\mu} X-A_{\mu ab}[T^{a},T^{b},X]\,,\\
\Tr~&\equiv&~\displaystyle{\int}{\rm d}^{3}y\sqrt{\hat{g}}\,,
\ea
\ee
where  $\hat{g}$ is an arbitrary function of $x,y$ which can be identified as the determinant of  the internal space metric $\hat{g}_{\hat{\mu}\hat{\nu}}(x,y)$. Then the whole six-dimensional space has the fiber bundle structure: at each point in $x$-space (base), there exists a corresponding  internal $y$-space (fiber).

Now we recall the BPS equation:
\be
D^{+}_\mu X^{\mu} =i\textstyle{{1\over\,3\sqrt{g}\,}}\e_{\mu\nu\lambda}[X^\mu, X^\nu, X^\lambda]\,.
\label{recallBPS}
\ee
Provided the above Nambu-bracket realization of the three-algebra, this BPS equation reads
\be
D^{+}_\mu X^{\mu} =i\textstyle{{1\over\,3\sqrt{g}\sqrt{\hat{g}}\,}}\e_{\mu\nu\lambda}
e^{\hat{\mu}\hat{\nu}\hat{\lambda}}\partial_{\hat{\mu}}X^\mu\partial_{\hat{\nu}}
X^\nu\partial_{\hat{\lambda}}X^\lambda=i\textstyle{{2\over\sqrt{\hat{g}}\,}}
\sqrt{\det\!\left(\partial_{\hm}X^{\l}\partial_{\hn}X_{\l}\right)}\,,
\ee
where the last equality holds since $\partial_{\hm}X^{\m}$ is a $3\times 3$ matrix. We integrate this formula  over $y$-space or take the trace. The final expression then leads to the usual Gauss law in three-dimension:
\be
\nabla\cdot E(x)=i\rho(x)\,,
\label{Erho}
\ee
with
\be
\ba{l}
E_{\mu}=\half\displaystyle{\int{\rm d}^{3}y}\,\sqrt{\hat{g}}\,X_{\mu}\,,\\
\rho=\textstyle{{1\over\,6\sqrt{g}\,}}\e_{\mu\nu\lambda}\displaystyle{\int}{\rm d}^{3}y~
\partial_{\hat{\mu}}\left(e^{\hat{\mu}\hat{\nu}\hat{\lambda}}X^\mu\partial_{\hat{\nu}}
X^\nu\partial_{\hat{\lambda}}X^\lambda\right)
=\displaystyle{\int}{\rm d}^{3}y~
\sqrt{\det\!\left(\partial_{\hm}X^{\l}\partial_{\hn}X_{\l}\right)}\,.
\ea
\ee
Remarkably, the density $\rho(x)$ matches with the Nambu-Goto action having the  $y$-space and the ``$X^{\mu}$-space" as the  world-volume and the target space.  Reflecting upon  the original untwisted BLG description of multiple M2-branes,
$X^{\mu}$ corresponds to three transverse scalars  and $x^{\mu}$ can be identified as three longitudinal physical directions in the  static gauge. In our twisted theory with three-algebra realized by Nambu-bracket, it is then natural to regard the  $(x,y)$-space and the $(x,X)$-space as the world-volume and the physical longitudinal space of an Euclidean M5-brane respectively with the partial static gauge ``$x=x$". This M5-brane picture then reveals that any point-like    BPS configuration in $x$-space or instanton may expand over $X$-space and in fact it corresponds to a Euclidean M2-brane.  The  $x$-space charge density $\rho(x)$ then measures the  volume of a Euclidean M2-brane in the $X$-space. Furthermore, $\rho$ being a surface integral,
if the $y$-space is  compact,  up to a $x$-space local  factor, the density $\rho(x)$ counts the winding number of
M2-branes wrapping three-cycles inside M5-brane. For a non-compact $y$-space, with a suitable boundary condition,
the integral may not vanish too.

Since the $Q$-transformations involve the three-commutators and depend on the $y$-space metric, our twisted Bagger-Lambert-Gustavsson theory is  topological only over the $x$-space but not over the $y$-space.

Furthermore with a six-dimensional metric:
\be
{\rm d}s^{2}_{6}=g_{\m\n}{\rm d}x^{\m}{\rm d}x^{\n}+\hat{g}_{\hm\hn}{\rm d}x^{\hm}{\rm d}x^{\hn}\,,
\ee
if we define  a $``(2,0)"$ and a $``(0,2)"$ two-form, respectively:
\be
\ba{ll}
B_{\mu\nu}:=\invsg\epsilon_{\mu\nu\lambda}X^{\lambda}\,,~~~&~~~
B_{\hat{\mu}\hat{\nu}}:=\textstyle{{1\over\,3\sqrt{g}\,}}\e_{\mu\nu\lambda}\, \left(
\partial_{\hat{\mu}}X^\mu\partial_{\hat{\nu}}
X^\nu-\partial_{\hat{\nu}}X^\mu\partial_{\hat{\mu}}X^\nu\right)X^\lambda\,,
\ea
\ee
then in terms of their  three-form field strengths,
\be
\ba{l}
H_{\lambda\mu\nu}:=D^{+}_{\lambda}B_{\mu\nu}+D^{+}_{\mu}B_{\nu\lambda}+D^{+}_{\nu}B_{\lambda\mu}=\invsg\epsilon_{\lambda\mu\nu}
D^{+}_{\rho}X^{\rho}\,,\\
H_{\hl\hm\hn}:=\nabla_{\hl}B_{\hm\hn}+\nabla_{\hm}B_{\hn\hl}+\nabla_{\hn}B_{\hl\hm}=\textstyle{\frac{1}{3\sqrt{\hat{g}}\sqrt{g}\,}}
\epsilon_{\hl\hm\hn}\epsilon_{\l\m\n}[X^{\l},X^{\m},X^{\n}]\,,
\ea
\ee
the BPS equation (\ref{recallBPS}) can be written in a compact form:
\be
H_{\lambda\mu\nu}=i\left(\ast H\right)_{\lambda\mu\nu}\,.
\ee
This corresponds to a partial self-duality equation of a three-form in Euclidean six dimension. It is partial, since it is the self-duality linking  $(3,0)$ and $(0,3)$ field strength and the other one linking $(2,1)$ and $(1,2)$ is missing.\footnote{It seems hard to find the $(1,1)$ two-form $B_{\mu\hm}$ which would complete the missing piece. One possible reason  might be that the scalar supercharges are not $\SO(6)$ chiral in contrast to the supersymmetries   of the six-dimensional M5-brane world-volume  theory~\cite{Howe:1983fr,Park:2001ut}.}

Provided these dictionaries  (despite of the incompleteness of the self-duality),  the BPS equation (\ref{recallBPS}) or (\ref{Erho})  indeed realizes the coupling of the self-dual three-form to the M2-brane charge density.

\section{Outlook\label{seccon}}
We have constructed a topological version of the BLG theory
and took some preliminary steps to study its physical contents.
But, clearly, more work would be required to reveal the
full physical contents of the topological theory.
First, an exhaustive list of observables should be found.
Second, a systematic study of the BPS configurations
and their contribution to the path integral should be done.
Finally, a perturbative computation of the partition function
and some of the observables should be carried out.
We hope to address these issues in a future work.

Another obvious direction is to consider other related theories.
In three dimensions, the minimum amount of supersymmetry needed to obtain a topological theory by twisting is $\CN=4$ \textit{i.e.~}eight supersymmetries.  An $\SO(3)$
subgroup of the $\SO(\CN)$ $R$-symmetry should be combined with the ``Lorentz'' $\SO(3)$ to yield a nilpotent, scalar supercharge $Q$
of the twisted theory.
But, for $\CN=3$, since the supercharge is a doublet
of Lorentz $\SO(3)$ and a triplet of the $R$-symmetry $\SO(3)$,
the twisting cannot
give rise to a scalar supercharge.

Recently, inspired by the BLG theory, a large class of $\CN\ge 4$
Chern-Simons theories (with ordinary Lie algebra gauge symmetry)
has been constructed \cite{CSmatter1,CSmatter2}
and their relation to string/${\cal M}$-theory has been elucidated.
It would be interesting to consider twisting those theories.
As they include both Chern-Simons terms as well as
non-linear sigma model with hyperK\"ahler target space,
they may reveal interesting connection between
the pure Chern-Simons theory \cite{purecs} and the Rozansky-Witten theory \cite{roz}.

\section*{Acknowledgments}
KL thanks Seungjoon Hyun for encouragement.
The work of KL is supported by the Basic Research Program of the Korea Science and
Engineering Foundation under grant number R01-2004-000-10651-0. SL is supported in part by the
KOSEF Grant R01-2006-000-10965-0 and the
Korea Research Foundation Grant KRF-2007-331-C00073.
The research  of JHP is supported in part  by   the Center for Quantum Spacetime of Sogang University with grant number R11 - 2005 - 021, and by the Korea Science and Engineering Foundation grant funded by the Korea government (R01-2007-000-20062-0).
\newpage

\appendix

\section{Some useful relations}
In curved backgrounds, the covariant derivative satisfies
\be
{}[D_{\mu},D_{\nu}]X^{\nu}-\tilde{F}_{\mu\nu}X^{\nu}+R_{\mu\nu}X^{\nu}=0\,,
\label{DDcom}
\ee
and hence
\be
\ba{ll}
\Tr\!\left(D_{\mu}X_{\nu},D^{\mu}X^{\nu}\right)=&
\half\Tr\!\left(D_{\mu}X_{\nu}-D_{\nu}X_{\mu},D^{\mu}X^{\nu}-D^{\nu}X^{\mu}\right)+
\Tr(D_{\mu}X^{\mu},D_{\mu}X^{\mu})\\
{}&+X^{\mu}_{a}\left(\tilde{F}_{\mu\nu}{}^{ab}X^{\nu}_{b}-R_{\mu\nu}X^{\nu a}\right)+
\partial_{\mu}\Tr(X^{\nu},D_{\nu}X^{\mu}) -\partial_{\mu}\Tr(X^{\mu},D_{\nu}X^{\nu})\,.
\ea
\ee
With this and the decomposition of the bosonic fields (\ref{forBoson}), up to total derivatives, we can rewrite a bosonic part of the Bagger-Lambert-Gustavsson action for the twist:
\be
\label{iden1}
\ba{ll}
\multicolumn{2}{l}{\half\Tr\!\left(D_{\mu}X_{I},D^{\mu}X^{I}\right)+\textstyle{\frac{1}{12}}
\Tr\!\left([X^{I},X^{J},X^{K}],[X_{I},X_{J},X_{K}]\right)}\\
~\equiv&
\textstyle{\frac{1}{4}}\Tr\!\left(D_{\mu}X_{\nu}-D_{\nu}X_{\mu}\,,\,D^{\mu}X^{\nu}-D^{\nu}X^{\mu}\right)
+\half \Tr\!\left(X^{\mu}, [D_{\mu},D_{\nu}]X^{\nu}\right)\\
{}&+\half\Tr\!\left(D_{\mu}X^{\mu}+i[X^{1},X^{2},X^{3}]\,,\,D_{\mu}X^{\mu}-i[X^{1},X^{2},X^{3}]\right)\\
{}&+\half\Tr\!\left(D_{\lambda}Y^{i}+i\half\epsilon_{\lambda\mu\nu}[X^{\mu},X^{\nu},Y^{i}],
D^{\lambda}Y_{i}-i\half\epsilon^{\lambda\rho\sigma}[X_{\rho},X_{\sigma},Y_{i}]\right)\\
{}&+\textstyle{\frac{1}{12}}\Tr\!\left([Y^{i},Y^{j},Y^{k}],[Y_{i},Y_{j},Y_{k}]\right)+
\textstyle{\frac{1}{4}}\Tr\!\left([Y^{i},Y^{j},X^{\mu}],[Y_{i},Y_{j},X_{\mu}]\right)\,.
\ea
\ee

Regarding the five-dimensional gamma matrices (\ref{fivegamma}),  two crucial  Fierz identities  follow
from the completeness relations of  $4\times 4$ symmetric and anti-symmetric matrices:
\be
\ba{l}
4\delta_{\alpha}{}^{\gamma}\delta_{\beta}{}^{\delta}+4\delta_{\beta}{}^{\gamma}\delta_{\alpha}{}^{\delta}
+(C\gamma^{ij})_{\alpha\beta}(\gamma_{ij}C^{-1})^{\gamma\delta}=0\,,\\
2\delta_{\alpha}{}^{\gamma}\delta_{\beta}{}^{\delta}-2\delta_{\beta}{}^{\gamma}\delta_{\alpha}{}^{\delta}
+C_{\alpha\beta}C^{-1}{}^{\gamma\delta}+(C\gamma^{i})_{\alpha\beta}(\gamma_{i}C^{-1})^{\gamma\delta}=0\,.
\ea
\label{5dFierz}
\ee
These further lead to other useful identities:
\be
\ba{rcl}
(\gamma^{i})^{\alpha}_{~\beta}(\gamma_{i})^{\gamma}_{~\delta} &=& 2\delta^{\gamma}_{~\beta}\delta^{\alpha}_{~\delta} -\delta^{\alpha}_{~\beta}\delta^{\gamma}_{~\delta}-2C^{-1 \alpha\gamma}C_{\beta\delta}\,,
\\
(\gamma^{i})^{\alpha}_{~\beta}(\gamma_{ji})^{\gamma}_{~\delta} &=&2\delta^{\alpha}_{~\delta}(\gamma_{j})^{\gamma}_{~\beta} -\delta^{\alpha}_{~\beta}(\gamma_{j})^{\gamma}_{~\delta}
-(\gamma_{j})^{\alpha}_{~\beta} \, \delta^{\gamma}_{~\delta}
-2(\gamma_{j}C^{-1})^{\alpha\gamma}C_{\beta\delta}\,,
\\
(\gamma^{i})^{\alpha}_{~\beta}(\gamma_{ijk})^{\gamma}_{~\delta}&=&
2\delta^{\gamma}_{~\beta}(\gamma_{jk})^{\alpha}_{~\delta}
-\delta^{\alpha}_{~\beta}(\gamma_{jk})^{\gamma}_{~\delta}-2C^{-1\alpha\gamma}(C\gamma_{jk})_{\beta\delta}
-(\gamma_j)^{\alpha}_{~\beta}(\gamma_{k})^{\gamma}_{~\delta}
+(\gamma_{k})^{\alpha}_{~\beta}(\gamma_{j})^{\gamma}_{~\delta}\,,
\ea
\ee
\be
\ba{rcl}
\delta^{\beta}{}_\gamma(\gamma_{ij})^{\alpha}{}_{\delta} + \delta^{\alpha}{}_\gamma(\gamma_{ij})^{\beta}{}_{\delta}
-(\gamma_{ij})^{\beta}{}_{\gamma}\delta^{\alpha}{}_{\delta} - (\gamma_{ij})^{\alpha}{}_{\gamma}\delta^{\beta}{}_{\delta}+2 C_{\gamma\delta}(\gamma_{ij}C^{-1})^{\alpha\beta}\\
+(\gamma_j)^{\beta}{}_{\gamma}(\gamma_i)^{\alpha}{}_{\delta}
-(\gamma_i)^{\beta}{}_{\gamma}(\gamma_j)^{\alpha}{}_{\delta}
+(\gamma_j)^{\alpha}{}_{\gamma}(\gamma_i)^{\beta}{}_{\delta}
-(\gamma_i)^{\alpha}{}_{\gamma}(\gamma_{j})^{\beta}{}_{\delta}&=&0\,,\\
-\delta^{\alpha}_{~\beta}(\gamma_{j})^{\gamma}_{~\delta}
-(\gamma_{j})^{\alpha}_{~\beta}\delta^{\gamma}_{~\delta}
-(\gamma_{j}C^{-1})^{\alpha\gamma}C_{\beta\delta}
-(C^{-1})^{\alpha\gamma}(C\gamma_{j})_{\beta\delta}
+\delta^{\gamma}_{~\beta}(\gamma_{j})^{\alpha}_{~\delta}
+(\gamma_{j})^{\gamma}_{~\beta}\delta^{\alpha}_{~\delta}&=&0\,.
\ea
\ee


\end{document}